\documentclass[%
 aip,
 jmp,%
 amsmath,amssymb,
 reprint,%
]{revtex4-1}

\usepackage{graphicx}
\usepackage{dcolumn}
\usepackage{bm}
\usepackage{hyperref}
\usepackage{natbib}
\begin{document}


\title{Identification of short-term and long-term time scales in stock markets and effect of structural break}

\author{Ajit Mahata}
\email{ajitnonlinear@gmail.com}
\affiliation{Department of Physics, National Institute of Technology
  Sikkim, Ravangla, Sikkim-737139, India.}
%
\author{Debi Prasad Bal}
\email{debi@nitsikkim.ac.in}
\affiliation{Department of Humanities and Social Sciences, National Institute of Technology Sikkim, Ravangla, Sikkim-737139, India}

\author{Md.Nurujjaman}
\email{jaman\_nonlinear@yahoo.co.in}
\affiliation{Department of Physics, National Institute of Technology
  Sikkim, Ravangla, Sikkim-737139, India.}
%

\date{\today}

\begin{abstract}
The paper presents the comparative study of the nature of stock markets in short-term and long-term time scales with and without structural break in the stock data. Structural break point has been identified by applying Zivot and Andrews structural trend break model to break the original time series (TSO) into time series before structural break (TSB) and time series after structural break (TSA). The empirical mode decomposition based Hurst exponent and variance techniques have been applied to the TSO, TSB and TSA to identify the time scales in short-term and long-term from the decomposed intrinsic mode functions. We found that for TSO, TSB and TSA the short-term time scales and long-term time scales are within the range of few days to 3 months and greater than 5 months respectively, which indicates that the short-term and long-term time scales are present in the stock market. The Hurst exponent is $\sim 0.5$ and $\geq 0.75$ for TSO, TSB and TSA in short-term and long-term respectively, which indicates that the market is random in short-term and strongly correlated in long-term. The identification of time scales at short-term and long-term investment horizon will be useful for investors to design investment and trading strategies.  

\end{abstract}

\maketitle

\begin{quotation}
Investors adopt different strategies in short-term and long-term depending on the investment time horizon as stock markets show different dynamics in different time scales. Hence identification of time scales in short-term and long-term is very important. The time scales have been identified using the empirical mode decomposition based Hurst exponent and normalised variance technique. The robustness of the analysis is further confirmed with Zivot and Andrews structural trend break model. Stock markets show random nature in short-term with time scales ranging from few days to nearly 3 months, and in long-term it shows long-range correlation with time scales greater than 5 months. We hope that these results may help the investors to take better decision on devising short-term and long-term trading strategies.
\end{quotation}

\section{\label{sec:intro}Introduction}
Stock market is a nonlinear, nonstationary and complex system, whose dynamical behaviour is mainly governed by the participation of different types of investor as well as various external parameters, and the dynamical behaviour is reflected in the stock price movement~\cite{bhgt,mantegna1999introduction,bouchaud2003theory,sornette2017stock,huang2003applications}. For a long time it was believed that the price variation of stock market is generated by the random process, and subsequently random walk model was suggested to explain the movement of future price change that does not depend on the past price and is not related with economic and fundamental information~\cite{alexander1964price}. Based on the Random walk theory, Efficient market hypothesis (EMH) was proposed by Fama (1965), which says that current stock price reflect all the available public information, that equally known to all investors, and hence no one can continuously make profit due to efficient nature of market~\cite{fama1995random}. 

Contrary to this thought, Mandelbrot showed during the period of development phase of Random theory that stock market possesses long memory and that can be described by the fractional Brownian motion~\cite{mandelbrot1963bb}, and subsequently several studies also suggest that market shows long-range memory~\cite{mantegna1999introduction,di2007multi,lin2013long,stanley2008statistical,kristoufek2013measuring,jung2008volatility,couillard2005comment,caporale2016long,1904.09214}, and can't be described fully by the EMH theory~\cite{leroy1976efficient,malkiel2003efficient}. 

Survey results of the fund managers and exchange dealers of different countries and various other works show that the investors segregate market in two main investment horizons in terms of time scales, namely short-term investment horizon (STIH) where technical analysis is applied and long-term investment horizon (LTIH) where fundamental analysis is used~\cite{menkhoff2010use,lui1998use}. Here time scales for STIH and LTIH are chosen empirically. Further it has been observed that the stock price movement is different in different time scales due to the different reaction and trading strategy of investors as well as various external parameters~\cite{conrad1998anatomy,jegadeesh1993returns,patzelt2018universal,mahata2019time}. Hence, identification of time scales for STIH and LTIH is very important for implementation of investment strategy.

Recently, in Ref.~\cite{mahata2019time} it is shown that the time scales for STIH and LTIH can be separated using empirical mode decomposition based Hurst exponent (EMDH) analysis, where, entire time series has been used for the analysis. However, it has been observed that structural change may happen in the stock data due to various factor~\cite{narayan2005oecd,chaudhuri2003random,worthington2007gold}, and hence, analysing the stock data considering the structural break is very important.  

In this paper we present the comparative study of the nature of stock markets in short-term and long-term investment horizon. Zivot and Andrews structural trend break model, empirical mode decomposition based Hurst exponent and variance techniques were applied to identify the short-term and long-term time scales, and  we found the time scales for short-term and long-term are within the range of few days to 3 months and greater than 5 months respectively. The identified time scales show that the short-term and long-term time scales are present in the stock markets. Using Hurst exponent analysis we found that markets is random in short-term and strongly correlated in long-term.

The rest of the paper is organised as follows: Sec~\ref{sec:MOA} discusses the method of analysis, Sec~\ref{sec:FDA} presents stock data analysed. Sec.\ref{sec:RAD} talk about results and discussion and finally Sec.\ref{sec:CON} provides the conclusions.    

\section{\label{sec:MOA}Method of analysis}
We have used Zivot and Andrews (ZA) (1992) structural trend break model-C to identify the structural break in the stock index time series~\cite{zivot2002further} as given below: 
\begin{equation}
\Delta X_t=c+\alpha X_{t-1}+\beta t+\theta D U_t+\gamma DT_t\displaystyle \sum_{j=1}^{k}d_j \Delta X_{t-j}+\epsilon_t,
\label{eqn:brk}
\end{equation} where $\Delta$ is the first difference of the variable, $DU_t$ is an indicator of dummy variable for a mean shift occurring at time trend break ($T_B$), where $1<T_B<T$. $DT_t$ is the corresponding trend shift variable, $\epsilon_t$ is white noise disturbance term and k is extra regressor added to eliminate possible nuisance-parameter dependencies in the limit distributions of the test statistics caused by temporal dependence in the disturbances.
$DU_t$ and $DT_t$ are defined as follows  

\[
DU_t=
\begin{cases}
1, &\text{if $t > T_B$;}\\
0, & \text{Otherwise}
\end{cases}
\]   and

\[
DT_t=
\begin{cases}
t-T_B, &\text{if $t > T_B$;}\\
0, & \text{Otherwise.}
\end{cases}
\].

The null hypothesis of Eqn.~\ref{eqn:brk} is $\alpha=0$. This indicates that the time series have unit root with a drift that excludes any structural break point. Whereas, the alternative hypothesis is $\alpha<0$ represents the time series is a trend stationary with one structural break point. By using ZA test we divided the original time series (TSO) of stock data into time series before structural break (TSB) and time series after structural break (TSA). Empirical mode decomposition (EMD) technique has been applied on the TSO, TSB and TSA to identify time scales in short-term and long-term. 

EMD technique is applied to decompose a signal where both nonlinearity and nonstationarity of the signal remain preserved~\cite{huang1998empirical,huang1998engineering}. The main advantage of the EMD technique is that it decomposes a signal into a numbers of {\it{intrinsic mode functions}} (IMF) of different time scales. An IMF satisfies the following two necessary conditions: (a) the numbers of extrema and zero crossings in the signal will be equal or differ by one; (b) the mean value of the envelope defined by the local maxima and the envelope defined by the local minima is zero. Detailed algorithm of calculating IMF are given below:

\begin{enumerate}
\item Lower envelope $U(t)$ and upper envelope $V(t)$ are drawn by connecting minima and maxima of the data respectively using spline fitting. 
\item Mean value of the envelope $m=[U(t)+V(t)]/2$ is subtracted from the original time series to get new data set $h= X(t)-m.$
\item Repeat the process (a) and (b) by considering $h$ as a new data set until the IMF conditions [(a) and (b)] are satisfied. Once the conditions are satisfied, the process terminates and $h$ is stored as IMF1. The IMF2 is calculated repeating the above steps (1)-(3) from the data set $d(t)=X(t)-IMF1$. When the final residual is monotonic in nature, the steps (a)-(c) are terminated and the orignal time series can be written as a set of IMFs plus trend, $X(t)=\displaystyle \sum_{i=1}^n IMF_i+residual,$ where $IMF_i$ represents the $i^{th}$ IMF. 
\end{enumerate}

The instantaneous frequency of each IMF can be estimated using the Hilbert transform define as $\displaystyle Y(t)=\frac{P}{\pi}\int_{-\infty}^{\infty}\frac{IMF(t)}{t-t'}dt$, where $P$ is the Cauchy principle value. The frequency can be calculated by $\omega=\frac{d\theta(t)}{dt}$, where, $ \displaystyle \theta(t)=tan^{-1}\frac{Y(t)}{IMF(t)}$~\cite{huang1998empirical}. As each IMF represents the mono-frequency component, the time scale of a particular IMF can be obtained from $\tau=1/\omega$.

Each IMF represents a signal with particular time scale. The IMF1 contains the lowest time scale present in the time series, and the IMF2 contains the second lowest time scale and so on. It can be concluded that the IMF1 fluctuates faster than the second and so on. Hence, EMD technique can be used to separate various important time scales present in a signal in the form of IMFs. As the main purpose of this work is to decompose the stock market indices with different time scales, IMFs will be very useful for this analysis. Identification of important IMF is very essential to separate the market dynamics in terms of short-term and long-term trading activities, and the identification of important IMF(s) can be done by using Hurst exponent ($H$) and Normalised variance ($NV$) analyses that are given below.    

Rescaled-range (R/S) analysis technique has been applied to find $H$ exponent of a time series~\cite{hurst1951long, mandelbrot1969robustness,mandelbrot1972statistical,mandelbrot1979robust}. Let us take a time series $X=\lbrace X_i, i=1,2,3,\cdot\cdot\cdot,N\rbrace$ of length N and divided the series into P subperiod of length n in such a way that $P\times~n=N$. Let us define each subperiod as $V_m$, where $m=1,2,3,\cdot\cdot\cdot,P$ and each element is $N_{q,m}$, where $q=1,2,3,\cdot\cdot\cdot,n$. Average and the standard deviation of each subperiod are expressed as 

$$\mu_m=\frac{1}{n}\displaystyle \sum_{q=1}^n{N_{q,m}}$$ and
$$S_m=\sqrt{\frac{1}{n}\displaystyle \sum_{q=1}^n(N_{q,m}-\mu_m)^2}$$ respectively. The cumulative departure of the time series is given by $Y_{q,m}=\displaystyle \sum_{j=1}^q(N_{j,m}-\mu_m)$, where $q=1,2,3,\cdot\cdot\cdot,n$. Range of the cumulative departure from the mean can be estimated by using $R_{V_m}=max(Y_{q,m})-min(Y_{q,m})$, where $1\leqslant q\leqslant n$. Individual subperiod range can be rescaled or normalized by dividing the standard deviation. So, rescaled range (R/S) can be represented as
$$R/S_n=\frac{1}{P}\displaystyle \sum_{m=1}^PR_{V_m}/S_{V_m}$$. 
In general, $(R/S_n)=c\times n^H$, where c is the constant and $H$ is the Hurst exponent. $H$ can be estimated from the slope of $log(R/S_n)$ versus $log(n)$. For a random time series, $H$ is 0.5, and for persistent and anti-persistent data, H is greater than 0.5, and less than 0.5 respectively.

$NV$ is an important statistical tool to analyse financial data that can be applied to separate signals with respect to their energy. It has been observed that the energy of a time series is mostly concentrated in its important IMF(s)~\cite{chatlani2012emd,zao2014speech}. The concentration of energy can be quantified by estimating the normalised variance $NV$ of the IMFs, and hence $NV$ can be used to identify the important IMFs. The $NV$ of an IMF can be calculated using the formula 
$$\displaystyle NV_i=\frac{\sqrt{\displaystyle \sum_t IMF_i^2(t)}}{\displaystyle \sum_{i=1}^N\sqrt{\sum_t IMF_i^2(t)}},$$ where, $N$ is the total number of IMF.

\section{\label{sec:FDA}stock data analysed}
\begin{figure}
\includegraphics[angle=0, width=8.5cm]{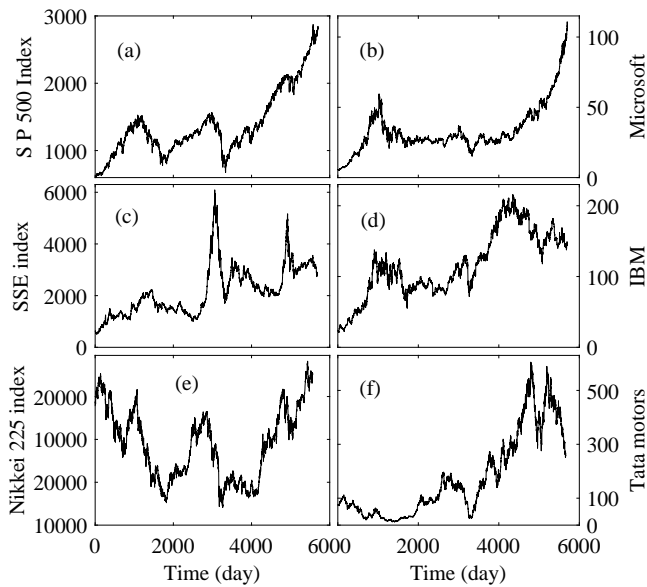}
\caption{\label{fig:raw_data} Original time series: (a) S\&P 500 (USA); (b) Microsoft (USA), (c) SSE (China); (d) IBM (USA); (e) Nikkei 225, (Japan); (f) Tatamotors (India).}
\end{figure}

In the present study, we have analysed the different stock indices and stock price of different countries such as (1) S\&P 500 (USA), (2) Nikkei 225 (Japan), (3) CAC 40 (France), (4) IBEX 35 (Spain) (5) HSI (Hong Kong), (6) SSE (China), (7) BSE SENSEX (India), (8) IBOVESPA (Brazil), (9) BEL 20 (Euro-Next Brussels), (10) IPC (Mexico), (11) Russel2000 (London),(12) TA125 (Israel) and companies like (13) IBM (USA), (14) Microsoft (USA), (15) Tata Motors (India), (16) Reliance Communication (RCOM) (India), (17) Apple (USA), (18) Reliance Industries Limited (RIL) (India). All the above data are taken from December 1995 to July 2018, and downloaded from yahoo finance~\cite{yahoo}. We have chosen the above indices and company mainly because of they are the leading global index. Figs.~\ref{fig:raw_data} show the stock indices of (a) S\&P 500 index (USA), (c) SSE Index (china), (e) Nikkei 225 index (Japan), and companies (b) Microsoft (USA), (d) IBM (USA) and (f) Tata Motors (India). The analysis of the data is given in the next section. 

\begin{figure*}
\includegraphics[angle=0, width=16 cm]{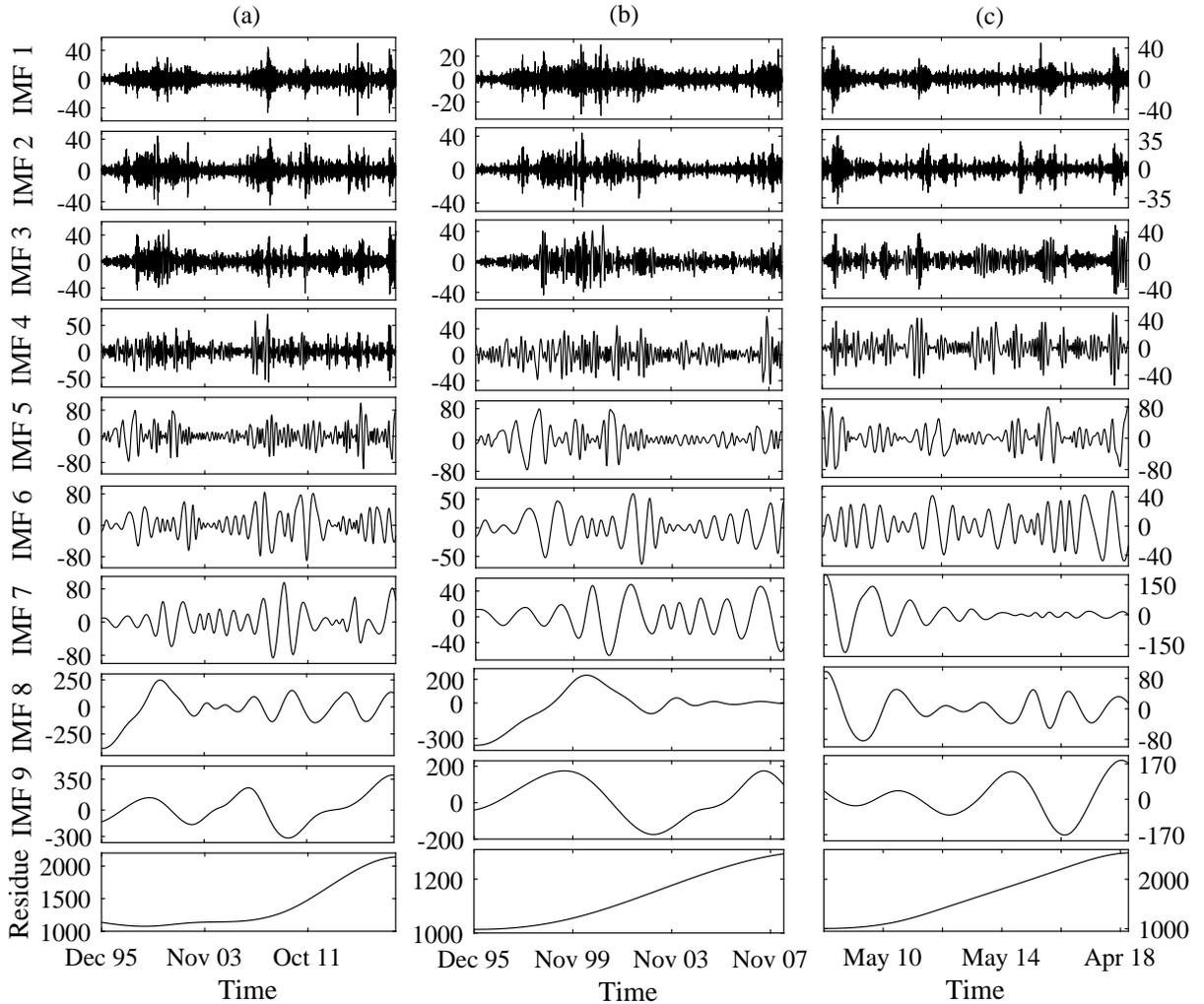}
\caption{\label{fig:SP_IMF} $(a)$, $(b)$ and $(c)$ are the IMFs obtained through empirical mode decomposition of the original time series, time series before and after structural break of the S\&P 500 index which is shown in Fig~\ref{fig:raw_data}(a). IMF1 represents the lowest time scale($\tau$) among all the IMFs, and $\tau$ increases with the IMF numbers. Residue represents the overall trend of the S\&P 500 index.}
\end{figure*}

\begin{figure}
\includegraphics[angle=0, width=8.5cm]{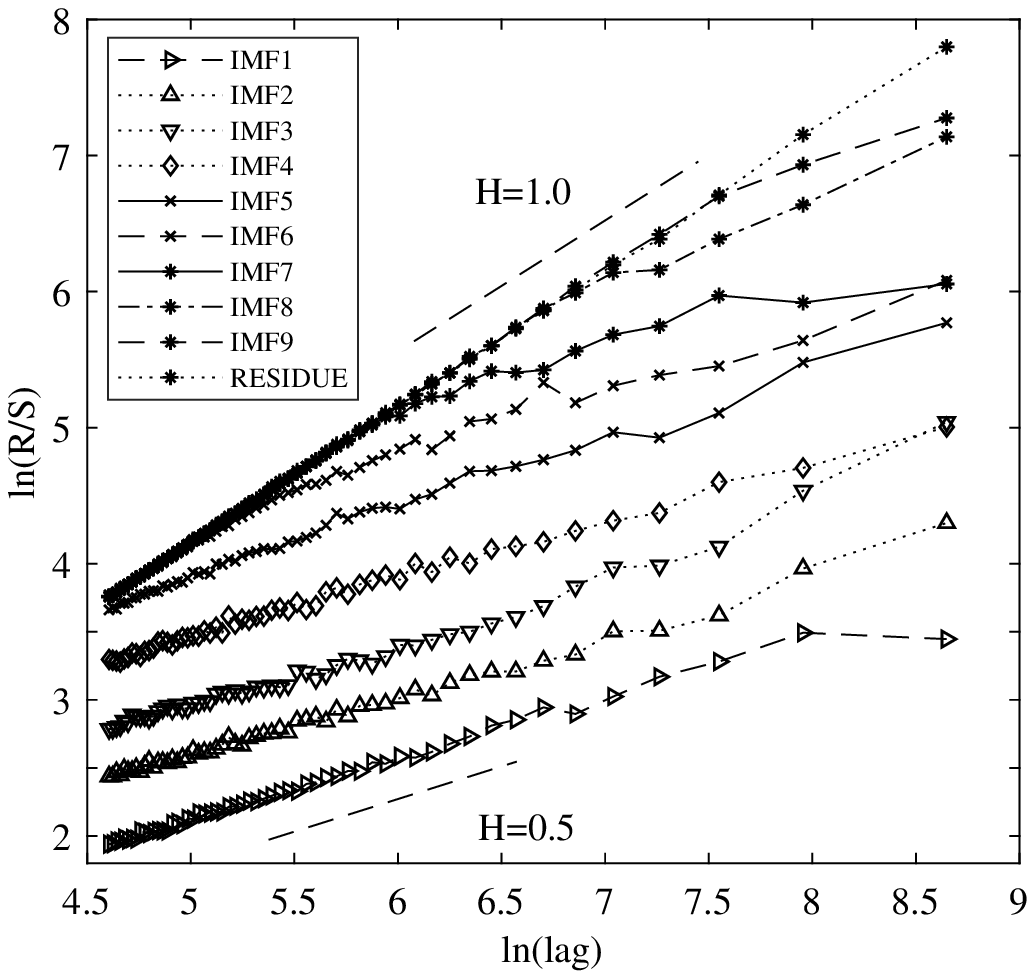}
\caption{\label{fig:Hurst_cal} $\ln(R/S)$ vs. $\ln(lag)$ plot for all the IMFs of S\&P 500 index. $H$ can be calculated from the slope.}
\end{figure}

\section{\label{sec:RAD}Results and Discussion}
In order to understand the behaviour of stock markets and stock price of various indices and companies at different time scales, we have applied the empirical mode decomposition (EMD) technique to decompose the original time series (TSO), time series before structural break (TSB) and time series after structural break (TSA) of stock index and price into IMFs of different time scales that is presented in the Subsec.~\ref{subsec:IMF}. 

\begin{figure}
\includegraphics[angle=0, width=8.5cm]{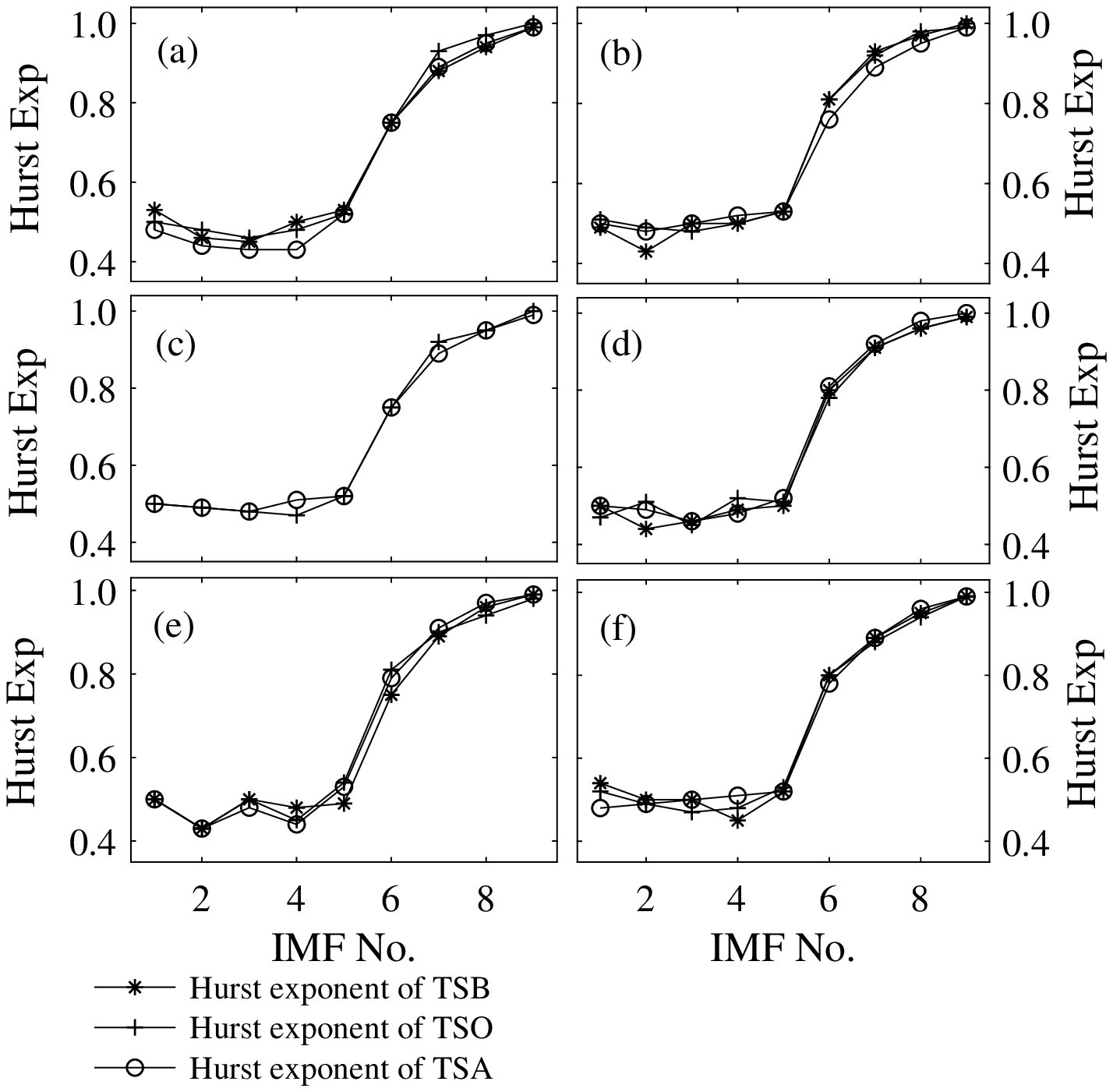}
\caption{\label{fig:Hurst_1-6} (a)-(f) represent the $H$ of S\&P 500, Nikkei 225, CAC 40 respectively. Time scales for S\&P 500; (a) $\tau=77D$, (b) $\tau=78D$ and (c) $\tau=65D$, Nikkei 225; (a) $\tau=66D$, (b) $\tau=76D$ and (c) $\tau=76D$, CAC 40;(b) $\tau=80D$ and (c) $\tau=78D$, IBEX 35;(a) $\tau=65D$, (b) $\tau=81D$ and (c)$\tau=81D$, HSI; (a) $\tau=65D$, (b) $\tau=97D$ and (c)$\tau=86D$, and SSE are (a) $\tau=91D$, (b) $\tau=92D$ and (c)$\tau=86D$ respectively for IMF1 to IMF5. (a), (b) and (c) represent the time scales of TSB, TSO and TSA respectively. The Figs.~\ref{fig:Hurst_1-6} show that the values of $H$ increases from IMF6. $H$ is nearly $0.5$ for IMF1 to IMF5, and $H$ value increases significantly for IMF6 to IMF9. Time scale unit D represents day.}
\end{figure}

\subsection{\label{subsec:IMF} EMD analysis}
Fig.~\ref{fig:SP_IMF} (a), (b) and (c) show nine intrinsic mode functions (IMFs) along with the residue separately of TSO, TSB and TSA of the S\&P 500 index respectively. Each IMF portrays the simple mono-frequency component of the stock data. All the IMF1 in Fig.~\ref{fig:SP_IMF} (a), (b) and (c) represents the mode with highest frequency, and the frequency gradually decreases with the increase of IMF numbers for TSO, TSB and TSA. Figs.~\ref{fig:SP_IMF} (a)-(c) show that the IMFs obtained from TSO [\ref{fig:SP_IMF} (a)], TSB [\ref{fig:SP_IMF} (b)] and TSA [\ref{fig:SP_IMF} (c)] are similar in nature. The last plots in Fig.~\ref{fig:SP_IMF} (a), (b) and (c) are the residue of the original time series, time series before and after structural break. The residue are the representation of the average trend of original time series, time series before and after structural break of the S\&P 500 index. In order to understand the nature of all similar IMFs of original time series, time series before and after structural break, detailed analysis using Hurst technique have been carried out in Subsec.~\ref{subsec:H}. 

\subsection{\label{subsec:H} Hurst exponent analysis}
\begin{figure}
\includegraphics[angle=0, width=8.5cm]{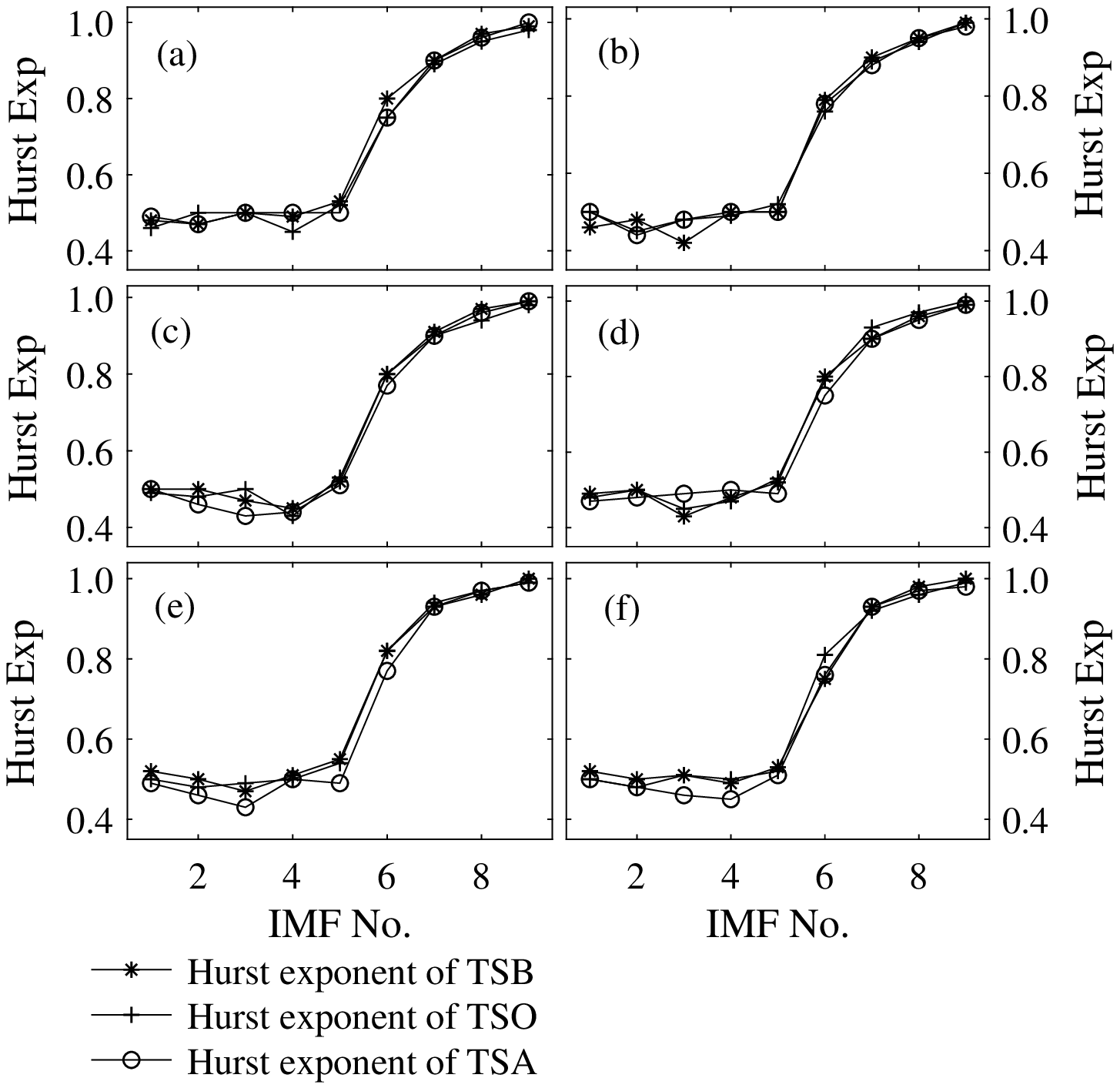}
\caption{\label{fig:Hurst_7-12} (a)-(f) represent the $H$ of BSE SENSEX, IBOVESPA, BELL 20, IPC, Russell 2000 and TA125 respectively. Time scales for BSE SENSEX; (a) $\tau=88D$, (b) $\tau=88D$ and (c)$\tau=57D$, IBOVESPA; (a) $\tau=57D$, (b) $\tau=80D$ and (c)$\tau=75D$, BELL 20; (a) $\tau=81D$, (b) $\tau=82D$ and (c)$\tau=79D$, IPC; (a) $\tau=73D$, (b) $\tau=85D$ and (c)$\tau=64D$, Russell 2000; (a) $\tau=91D$, (b) $\tau=94D$ and (c)$\tau=62D$, and TA125 are (a) $\tau=63D$, (b) $\tau=79D$ and (c)$\tau=60D$ respectively for IMF1 to IMF5. (a), (b) and (c) represent the time scales of TSB, TSO and TSA respectively. The Figs.~\ref{fig:Hurst_7-12} show that the values of $H$ increases from IMF6. $H$ is nearly $0.5$ for IMF1 to IMF5, and $H$ value increases significantly for IMF6 to IMF9. Time scale unit D represents day.}
\end{figure}

Hurst exponent ($H$) of all the decomposed IMFs has been calculated using rescaled-ranged analysis presented in Sec.~\ref{sec:MOA}. Fig~\ref{fig:Hurst_cal} shows the typical plots of $\ln(R/S)$ vs. $\ln(lag)$ of IMF1 to IMF9 and residue. Slope of the each plot is basically $H$ value of the corresponding IMF. From the Fig~\ref{fig:Hurst_cal} it is clear that slope of IMF1 to IMF5 is nearly 0.5 and suddenly increases. The value of $H\sim 1.0$ for residue of the index.  

\begin{figure}
\includegraphics[angle=0, width=8.5cm]{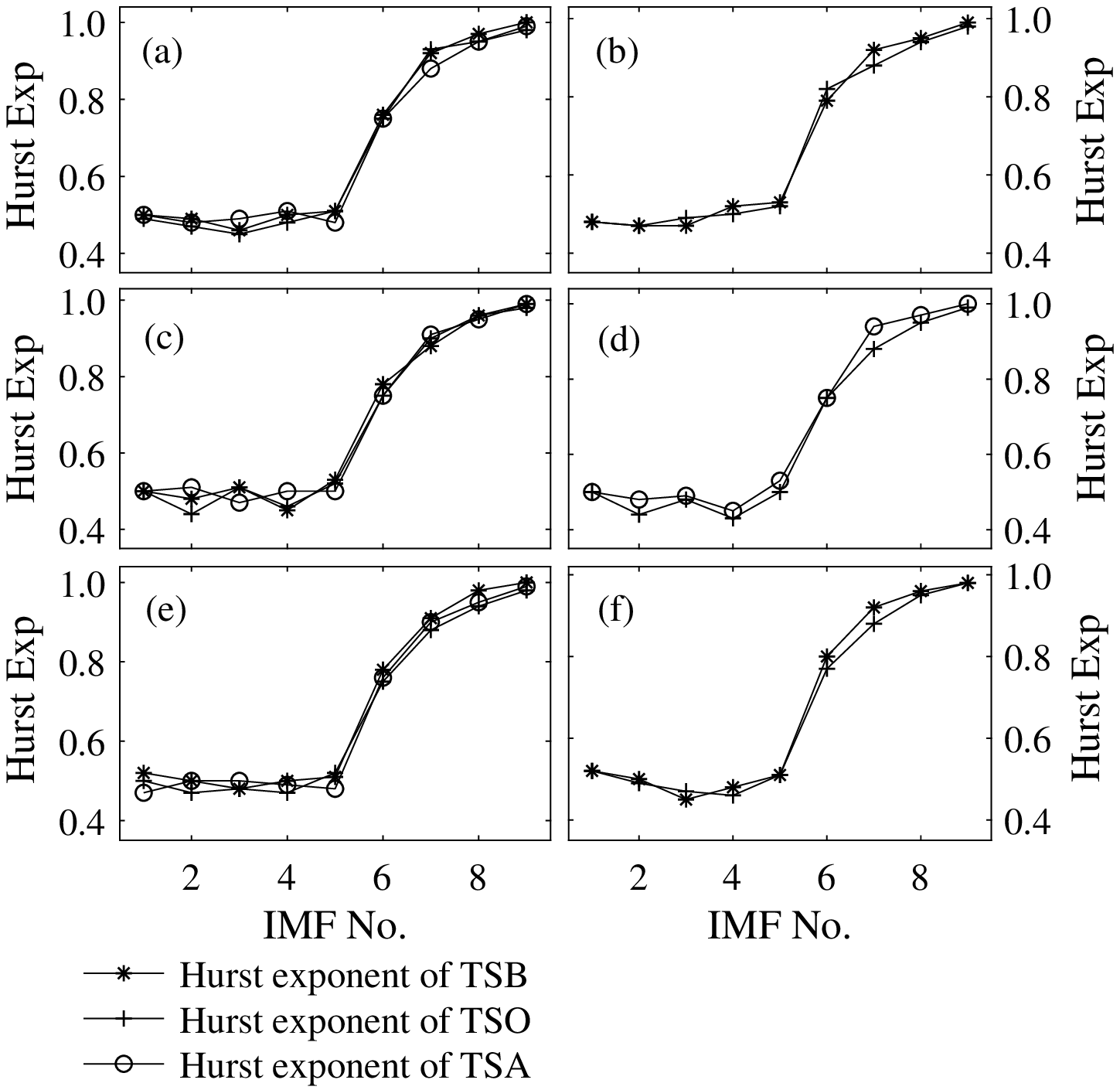}
\caption{\label{fig:Hurst_13-18} (a)-(f) represent the $H$ of IBM, Microsoft, Tatamotors, RCOM, Apple and RIL respectively. Time scales for IBM; (a) $\tau=86D$, (b) $\tau=93D$ and (c)$\tau=57D$, Microsoft; (a) $\tau=83D$ and (b) $\tau=86D$, Tatamotors; (a) $\tau=88D$, (b) $\tau=91D$ and (c)$\tau=65D$, RCOM; (b) $\tau=96D$ and (c)$\tau=70D$, Apple; (a) $\tau=76D$, (b) $\tau=76D$ and (c)$\tau=61D$, and RIL are (a) $\tau=77D$ and (b) $\tau=78D$ respectively for IMF1 to IMF5. (a), (b) and (c) represent the time scales of TSB, TSO and TSA respectively. The Figs.~\ref{fig:Hurst_13-18} show that the values of $H$ increase from IMF6. $H$ is nearly $0.5$ for IMF1 to IMF5, and $H$ value increases significantly for IMF6 to IMF9. Time scale unit D represents day.}
\end{figure}

Figs.~\ref{fig:Hurst_1-6} (a)-(f) show the value of $H$ of the IMF1 to IMF9 of the original time series, the time series before and after structural break of S\&P 500, Nikkei 225, CAC 40, IBX35, HSI, and SSE respectively. The plots show that the value of $H$ is around 0.5 for IMF1 to IMF5 of the original time series, the time series before and after structural break. The value of $H$ increases suddenly to $\sim0.75$ for IMF6 and gradually increases till the last IMF for the all three time series. The $H$ value for IMF6 to IMF9 is between 0.75 to 1.0. It is already shown that $H\sim 1$ for residue of all the data. Similarly, Figs.~\ref{fig:Hurst_7-12}(a)-(f) show the $H$ of the IMF's of BSE SENSEX, IBOVESPA, BELL 20, IPC, Russell 2000 and TA125 respectively, and Figs.~\ref{fig:Hurst_13-18}(a)-(f) show the $H$ of the IMF's of IBM, Microsoft, Tatamotors, RCOM, Apple and RIL respectively. It is pertinent to mention that we were not able to calculate $H$ exponent of CAC 40 for TSB, Microsoft for TSA and RCOM for TSB, RIL for TSA because of short data lengths. 

\begin{figure*}
\includegraphics[angle=0, width=16cm]{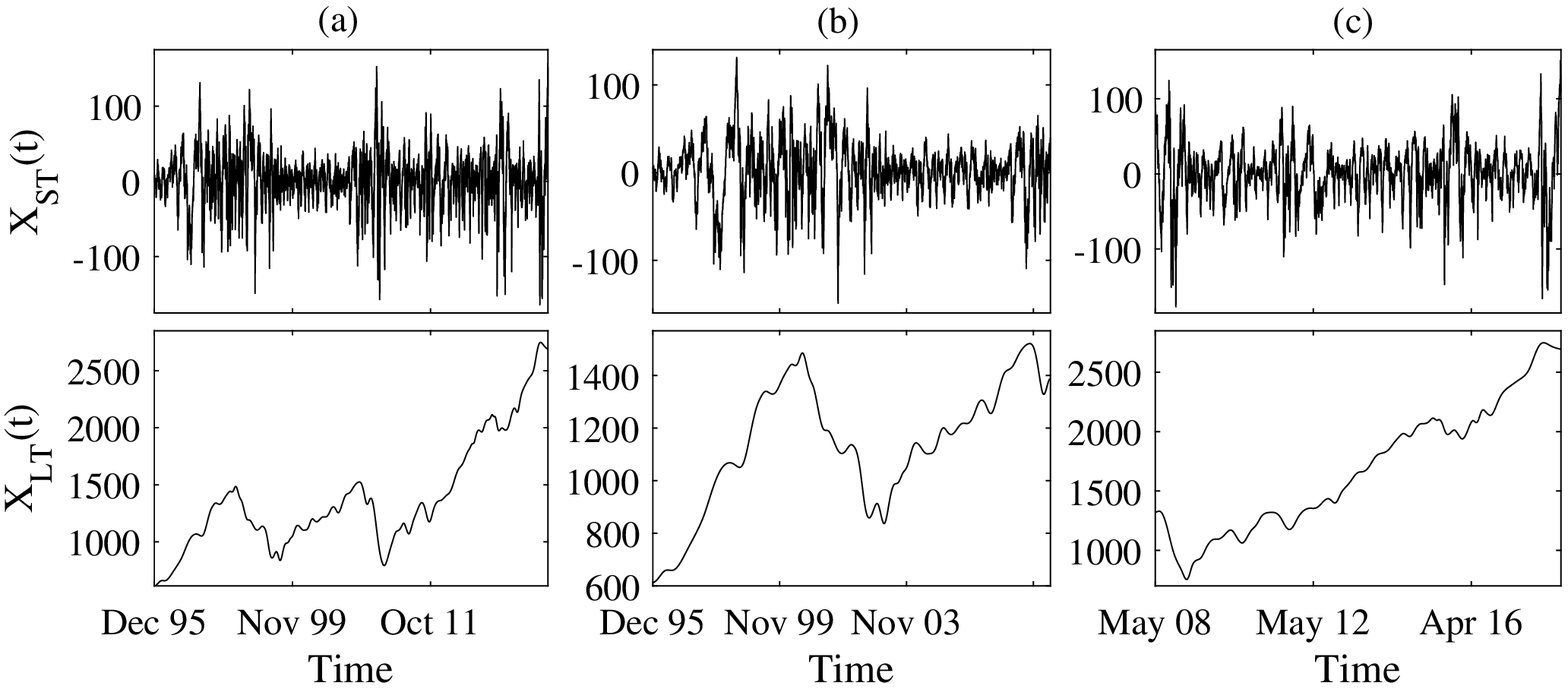}
\caption{\label{fig:ST_LT} (a), (b) and (c) represent the reconstructed short-term ($X_{ST}(t)$) and long-term ($X_{LT}(t)$) of TSO, TSB and TSA respectively of S\&P 500 index.}
\end{figure*}

From Figs.~\ref{fig:Hurst_1-6}, ~\ref{fig:Hurst_7-12} and ~\ref{fig:Hurst_13-18} it is clear that the value of $H$ is around 0.5 for IMF1 to IMF5 of all the indices and companies stock prices. We can conclude from the $H$ exponent values that the IMF1 to IMF5 are random in nature. Therefore, stock price represented by IMF1 to IMF5 is random. The $H$ value between 0.75 to 1 for IMF6 to IMF9 and residue indicates that they are having long-range correlation which in turn means less volatility and less roughness. Therefore, stock price represented by IMF6 to IMF9 and residue is persistent in nature.

\begin{figure}
\includegraphics[angle=0, width=8.5cm]{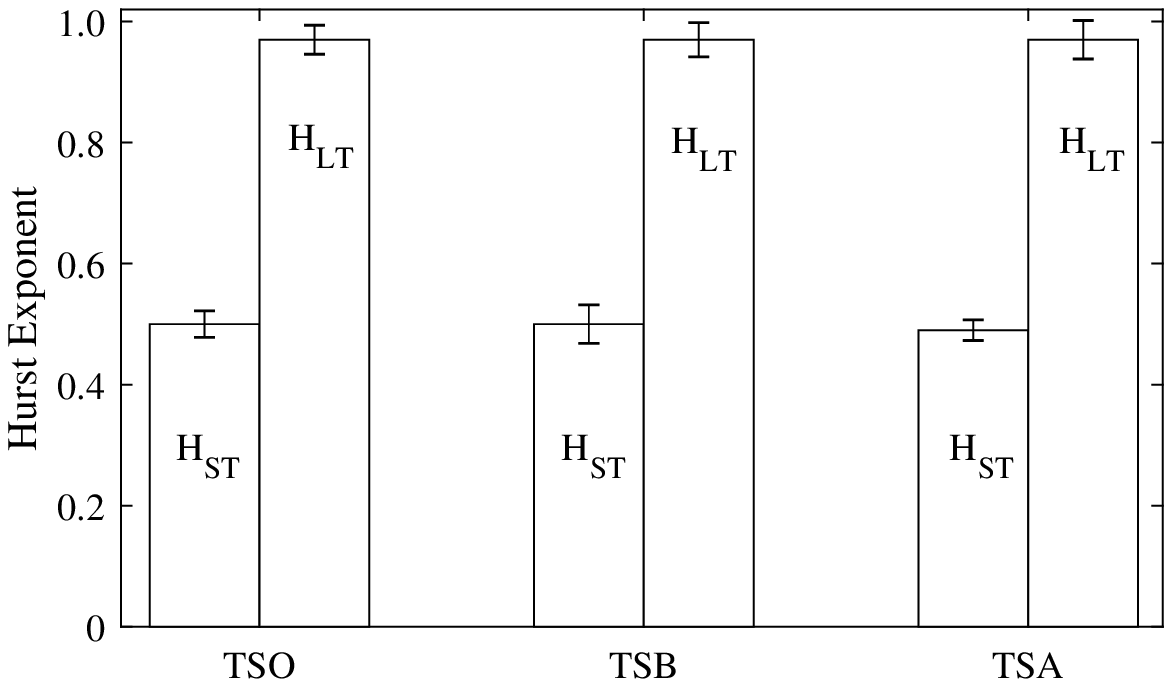}
\caption{\label{fig:RCONH} Represents the bar plot of the $H$ exponent with $2\sigma$ error bar for the TSO, TSB and TSA of all the indices and companies, where $H_{ST}$ and $H_{LT}$ represent the $H$ exponent of the reconstructed $X_{ST}(t)$ and $X_{LT}(t)$. $H_{ST}\sim0.5$ and $H_{LT}\sim 0.97$ for all the three cases.}
\end{figure}

In order to separate the dynamics, we have constructed two time series, $X_{ST}(t)=\displaystyle \sum_{i=1}^5~IMF_i$ adding IMF1 to IMF5 and $\displaystyle X_{LT}(t)=[\displaystyle \sum_{i=6}^9~IMF_i+Residue]$ for the original time series, the time series before and after structural break of all the stock index and data respectively. Figs.~\ref{fig:ST_LT} (a)-(c) shows the reconstructed short-term $X_{ST}(t)$ and long-term $X_{LT}(t)$ time series. Finally we have calculated the $H$ exponent of the reconstructed time series  $X_{ST}(t)$ and $X_{LT}(t)$ of three time series which is plotted in Fig~\ref{fig:ST_LT}. The plot ~\ref{fig:RCONH} shows that $H\sim 0.5$ for all reconstructed time series $X_{ST}(t)$, and $H\sim 0.97$ for $X_{LT}(t)$. The results show that two reconstructed time series $X_{ST}(t)$ and $X_{LT}(t)$ actually separates the random and long-range dynamics in the system. It is pertinent to mention that $X_{LT}(t)$ represents the collective fundamentals of the companies as stated in Ref.~\cite{mahata2019time}

\begin{figure*}
\includegraphics[angle=0, width=16cm]{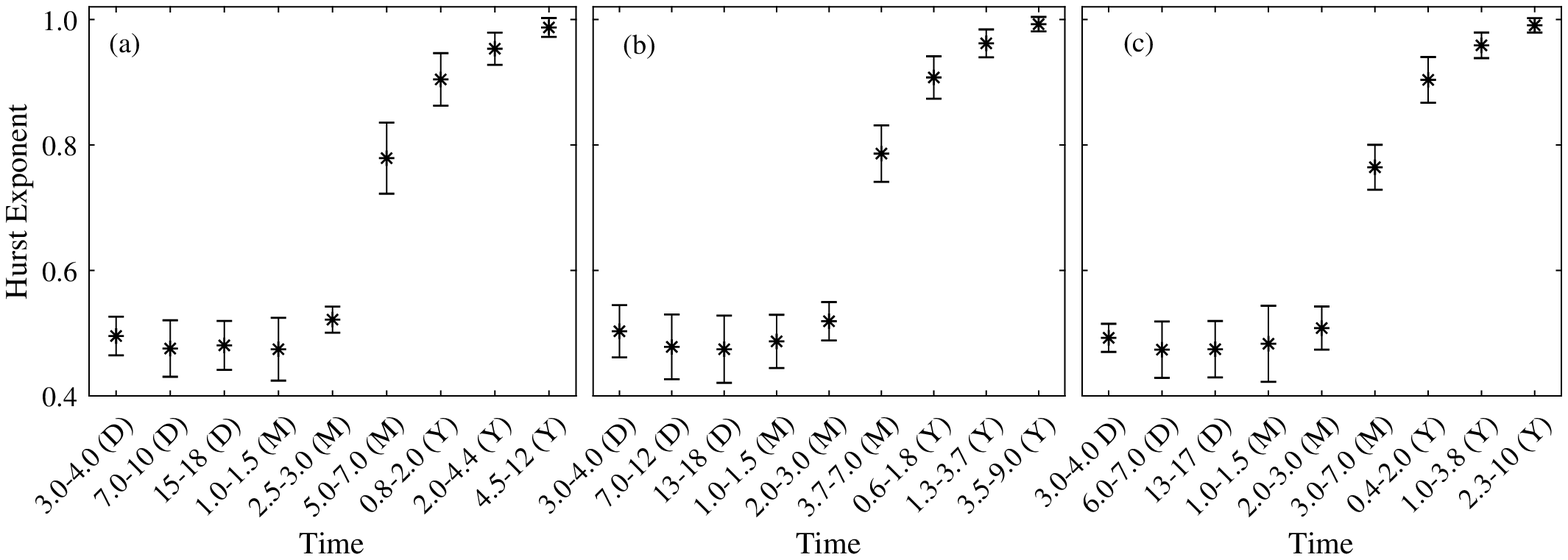}
\caption{\label{fig:Hurst_brk1-3} (a), (b) and (c) represent the $H$ of TSO, TSB and TSA respectively of all the indices and companies with $2\sigma$ error bar. The plots show that the $H\sim 0.5$ for the decomposed time series (IMF1 to IMF5) with time scales around 3 days to 3 months. The value of $H$ suddenly increases to $\sim 0.75$ with time scales 5-7 months and reaches approximately 1.0 with the increase in time scales for all three cases for IMF6 to IMF9.}
\end{figure*}

In order to identify the time scales in two types of dynamics, we have estimated the time scales of all the IMFs of the original time series, the time series before and after structural break of all the stock index and data. Fig.~\ref{fig:Hurst_brk1-3} (a), (b) and (c) show the time scales versus $H$ exponent for TSO, TSB and TSA respectively calculated from their IMFs. Fig.~\ref{fig:Hurst_brk1-3} (a), (b) and (c) show that the $H\sim 0.5$ for the time series with time scales few days to 3 months, i.e., for IMF1 to IMF5. Therefore, it may be concluded that the markets in ITH of few days to 3 months are random in nature. The value of $H$ suddenly increases to 0.75 for the time scales 5-7 months for TSO, 3.0-7 months for TSB and 3.7-7 months for TSA, from IMF6 as shown in the same figures, and with increase in time scales $H$ approaches 1. As $H>0.5$ indicates the long-range correlation, we can conclude that the stock markets in time scales greater than five months show long-range correlation. 

\subsection{\label{subsec:Var} Normalised variance analysis}
It has been observed that the energy of a time series is mostly concentrated in its important decomposed IMFs~\cite{chatlani2012emd,zao2014speech}, and the energy can be quantified by calculating the normalised variance ($NV$) of the IMFs that has already been described in Sec~\ref{sec:MOA}.

\begin{figure}
\includegraphics[angle=0, width=8.5cm]{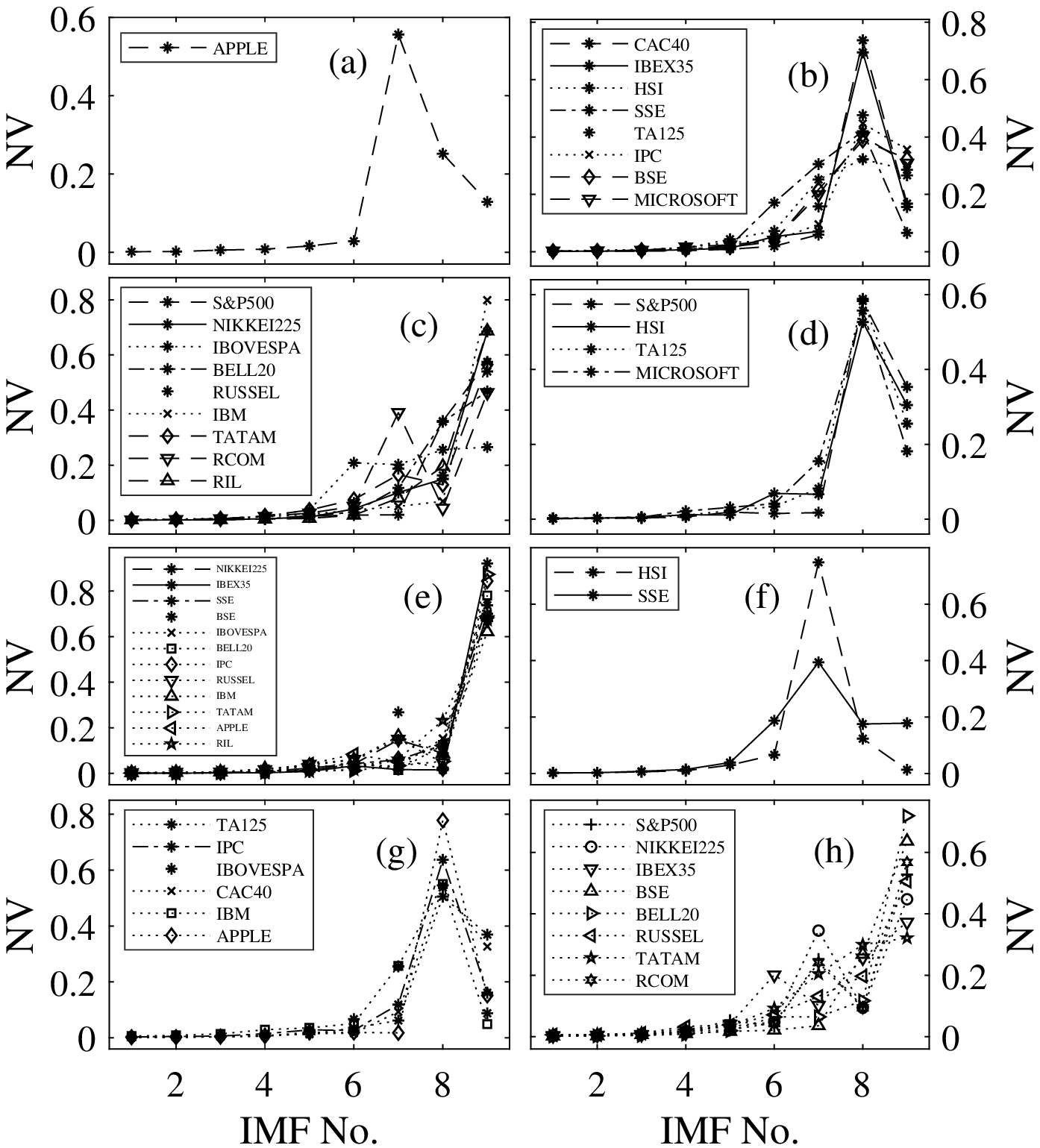}
\caption{\label{fig:Var1-3}(a)-(c), (d)-(e) and (f)-(h) represent the $NV$ of TSO, TSB and TSA respectively of all the indices and companies. Plot (a) shows IMF7 is important for Apple; Plot (b) shows IMF8 is important for several indices and companies shown in plot legends; and plot (c) shows that the IMF9 is important for other indices and companies shown in the legends. Similarly, Same kind of of results have been obtained  for TSB and TSA as shown in the plots (d)-(h).}
\end{figure}

Figs.~\ref{fig:Var1-3}(a)-(c), (d)-(e) and (f)-(h) show the normalised  variance ($NV$) of the IMF's of TSO, TSB and TSA of the stock indices and company price respectively. Fig.~\ref{fig:Var1-3}(a) shows IMF7 with time scale 2.5 months is important; Fig.~\ref{fig:Var1-3}(b) shows that IMF8 is important for several indices and companies as given in plot legends of the Figure, where time scale of the IMF is around 0.8 to 2 yrs. Fig.~\ref{fig:Var1-3}(c) shows that IMF9 is important for other indices and companies shown in the legends. Similar results have been obtained for TSB and TSA time series as shown in the plots (d)-(h). The results show that in long-term investment horizon, different time scales may be important depending on the market dynamics of particular index.

\section{\label{sec:CON}Conclusions}
Our study reveals that the stock markets possesses the long-range correlation and randomness depending on time scales. We have separated market dynamics into short-term and long-term by using empirical mode decomposition based Hurst exponent analysis. Decomposition has been done of each index and company into nine intrinsic mode functions (IMF) and a residue. Each of the IMFs have different time scales and frequency. We found that for IMF1 to IMF5 $H\sim 0.5$, and for IMF6 to IMF9 and residue $H\geq 0.75$. We have also obtained similar results for the TSO, TSB and TSA of all the stock index and data obtained by applying Zivot and Andrews structural trend break model. The results show that market is random with time scales from few days to 3 months and shows long-range correlation with time scales $\geq 5$ months.

Our finding gives an idea about the behavior of the stock markets which may be applicable for both the short-term and long-term investors. Our results suggest that short-term trading is very risky due to presence of randomness in shorter investment horizon. However, in long-term trading, risk is less because of the presence of long-range correlation in the stock markets.

\section*{Acknowledgment}
We would like to acknowledge Jean-Philippe Bouchaud for some valuable discussion and suggestions.
\bibliography{ref}
\end{document}